\documentclass[floats,floatfix,showpacs,amssymb,prd,twocolumn,superscriptaddress,nofootinbib]{revtex4-1}

\usepackage{graphicx,epsf, epsfig, amssymb,amsmath}
\usepackage{bm}
\usepackage{longtable}
\usepackage[usenames,dvipsnames]{xcolor}
\usepackage{color}
\usepackage[breaklinks]{hyperref}
\usepackage{amsfonts,amsmath,amssymb,mathrsfs}
\usepackage{multirow}
\usepackage[nolist,nohyperlinks]{acronym}
\usepackage{xspace}
\usepackage{booktabs,array}
\usepackage[british]{babel}
\usepackage[utf8]{inputenc}

\newcommand{\tn}{\textnormal}
\newcommand{\ok}{\omega_\tn{k}}

\begin{document}

\title{Relativistic tidal effects in nonstandard Kerr space-time}

\author{Andrea Maselli}
\email{andrea.maselli@uni-tuebingen.de}
\affiliation{Theoretical Astrophysics, Eberhard Karls University of Tuebingen, 
Tuebingen 72076, Germany}

\author{Kostas Kokkotas}
\email{kostas.kokkotas@uni-tuebingen.de}
\affiliation{Theoretical Astrophysics, Eberhard Karls University of Tuebingen, 
Tuebingen 72076, Germany}
\affiliation{Center for Relativistic Astrophysics and School of Physics,
Georgia Institute of Technology, Atlanta, Georgia 30332, USA}

\author{Pablo Laguna}
\email{plaguna@gatech.edu}
\affiliation{Center for Relativistic Astrophysics and School of Physics,
Georgia Institute of Technology, Atlanta, Georgia 30332, USA}


\begin{abstract}
Astrophysical phenomena involving massive black holes (BHs) in close binaries 
are expected to leave detectable signatures in the electromagnetic and gravitational-wave 
spectrum. Such imprints may provide precious information to probe the 
space-time around 
rotating BHs and to reveal new insights on the nature of gravity in the strong-field 
regime. To support this observational window, it is crucial to develop suitable tests
 to verify the predictions of General Relativity. In this framework, the metric 
recently proposed by Johannsen and Psaltis parametrizes strong-field deviations from 
a Kerr space-time in a theory-independent way. In the following, we make use of this 
approach to describe the tidal field produced by spinning BHs. We compute the 
gravito-magnetic and gravito-electric tidal tensors for particles moving on equatorial circular 
geodesics, comparing our results with those obtained in the standard General Relativity scenario.  
Our calculations show significant differences even for distances far form the 
last stable orbit, which may affect the evolution of the binary and leave detectable signatures. 
We test our framework computing quasiequilibrium sequences of BH-white dwarf systems by means 
of the affine model, for different binary configurations.
\end{abstract}

\maketitle


\section{Introduction}

Since its formulation 100 years ago, 
General Relativity (GR) has successfully passed a large set of observational and 
experimental tests \cite{Will:2014xja}. Most of them however, probed only the weak field 
regime of gravity, and therefore a number of strong-field GR predictions still remain to 
be verified \cite{Berti:2015itd,Yunes:2013dva,Psaltis:2008bb,Bambi:2015kza}.
Black holes are among the most genuine of such predictions, with no 
analog in the Newtonian theory, and represent the ideal candidates to test gravity 
under extreme conditions. In General Relativity, black holes (BHs) belong to the Kerr family, and 
according to the no-hair theorem, their exterior stationary and isolated gravitational field 
depend only on two parameters: their mass and angular momentum \cite{wiltshire2009kerr}. 
Initial deviations from the Kerr metric are rapidly radiated away by the emission of 
gravitational radiation \cite{PhysRevD.5.2419}. 
A proof of the validity of the no-hair theorem is still lacking. However, future 
electromagnetic \cite{2014ApJ...784....7B,2008SPIE.7013E..2AE,feroci2012large,gendreau2012neutron} 
and gravitational-wave \cite{ 2010CQGra..27h4006H,2008CQGra..25k4045A} observations 
promise to shed new light on this scenario, and are expected to prove the {\it Kerr hypothesis}.
In this regard, several efforts have been devoted to develop independent tests to determine 
the features of the strong gravitational field in the BH surroundings. Such tests follow a 
bottom-up approach, in which the BH space-time is parametrized in a phenomenological way, 
with the aim to map possible detected deviations in terms of an alternative theory of gravity. 
Requiring that the new metric is free of pathologies as naked singularities or closed timelike 
curves makes these studies an extremely difficult task. We refer the reader to Ref. 
\cite{PhysRevD.87.124017} and references therein, for a systematic study of the main 
features of some parametric frameworks which have been proposed in the literature. 

In this work, we make use of the new approach recently 
developed  by Johannsen and Psaltis (JP) \cite{PhysRevD.83.124015}. The authors introduced 
polynomial corrections into the Schwarzschild metric as initial seeds, showing that this ansatz 
can be turned into a Kerr-like metric through the Newmam-Janis algorithm 
\cite{:/content/aip/journal/jmp/6/6/10.1063/1.1704350}.  
The mathematical properties and the topology of the JP metric, as well as their 
astrophysical implications have been extensively studied in Refs. 
\cite{0004-637X-716-1-187,PhysRevD.83.124015,PhysRevD.87.124017,PhysRevD.83.124015,Bambi201113,PhysRevD.85.043002}.
Moreover, tests involving properties of iron lines, quasiperiodic oscillations, 
continuum spectra of accretion disks and images of the accretion flows have 
been analyzed in Refs. 
\cite{Bambi20115,PhysRevD.87.124010,1475-7516-2012-09-014,PhysRevD.87.023007,2013ApJ...773...57J,0004-637X-761-2-174}. This metric has been also extended to a more general 
parametrization in Ref. \cite{PhysRevD.89.064007}, where the authors addressed some 
unexplored key features of the original framework.

In this paper, we investigate the effects of strong-gravity corrections captured by the JP 
approach, on the tidal field produced by rotating BHs. We derive the expressions for the 
gravito-magnetic and gravito-electric tidal tensors, which act as source of the geodesic 
deviation of nearby test particles, and determine the frame-dragging precession of 
test gyroscopes. The results of this work can be useful to devise tests of GR 
through astrophysical observations of close binaries involving a massive 
BHs and a companion star. Such environments may lead to tidal disruption events 
even at large distances, producing detectable gravitational and x-ray/UV 
radiation \cite{0004-637X-615-2-855}.

In order to test our theoretical framework, we simulate the orbital evolution of BH-white 
dwarf (WD) binaries, using the new formulation of the gravito-electric tidal tensor together 
with the affine model, which is a semianalytical approach to describe star deformations 
induced by an external quadrupolar tidal field. Originally developed to study the evolution 
of normal stars or WDs within a Newtonian scheme 
\cite{1985MNRAS.212...57L,1985MNRAS.212...23C,0004-637X-532-1-530}, 
this model was recently improved to describe neutron star tidal disruption events 
in compact binaries, taking into account relativistic effects on the stellar structure 
\cite{0264-9381-26-12-125004,PhysRevD.81.064026}, and post-Newtonian 
corrections both on the orbital dynamics and the tidal field 
\cite{Ferrari:2011as,Maselli:2012zq}.

With this framework, we follow quasiequilibrium sequences of prototype BH-WD 
binaries for different modified Kerr metrics, finding that the stellar deformations 
may vary with respect to the GR case up to $5\%$ even for large distances. 
For each binary configuration, we also identify the onset of the mass transfer from 
the star to the companion object, which can be used as initial data for fully relativistic 
numerical simulations, to investigate the properties of the accreting flow onto the BH.

This paper is organised as follows. In Sec. \ref{sec:JPmetric}, we introduce 
the JP metric, and we derive the basic features of geodesic motion. 
In Sec. \ref{sec:tidalfield}, we describe the procedure to characterize the tidal
field in the JP space-time, and we explicitly compute the gravito-magnetic and 
gravito-electric tidal tensors. Moreover, we discuss the relevance 
of the corrections induced by the strong-gravity modifications of the Kerr metric, 
comparing our analytical results with those obtained for the pure GR scenario. 
In Sec. \ref{sec:numerical}, we numerically investigate tidal disruption events in 
BH-WD binaries, for different configurations. 
Finally, in Sec. \ref{sec:concl}, we draw the conclusions.

We use Greek letters $(\alpha,\beta,\ldots)$ to denote space-time indices 
and Latin characters $(i,j,\ldots)$ for spatial indices.

\section{JP metric}\label{sec:JPmetric}
The Johannsen-Psaltis metric is described in Boyer-Lindquist 
coordinates $x^\mu=(t,r,\theta,\phi)$ by the following line element, 
\begin{align}\label{JPmetric}
ds^2=&-(1+h)\left(1-\frac{2Mr}{\Sigma}\right)dt^2
+\frac{\Sigma(1+h)}{\Delta+a^2\sin^2\theta h}dr^2\nonumber\\
&+\Sigma d\theta^2-\frac{4aMr\sin^2\theta}{\Sigma}(1+h)dtd\phi\nonumber\\
&+\left[r^2+a^2+\frac{2a^2Mr\sin^2\theta}{\Sigma}+h\frac{a^2(\Sigma+2Mr)}{\Sigma}\sin^2\theta\right]\nonumber\\
&\times\sin^2\theta d\phi^2\ ,
\end{align}
where $\Sigma=r^2+a^2\cos^2\theta$ and $\Delta=r^2+a^2-2Mr$. 
The function $h(r,\theta)$ parametrizes the deviations from the {\it pure} 
Kerr space-time and is given by
\begin{equation}\label{JPh}
h(r,\theta)=\sum_{k=0}^{\infty}\left(\epsilon_{2k}+\epsilon_{2k+1}\frac{Mr}{\Sigma}\right)
\left(\frac{M^2}{\Sigma}\right)^k\ .
\end{equation}
The JP metric has an infinite number of deformations parameters. However, 
some of them are constrained by theoretical and experimental bounds. As 
noted in Ref. \cite{PhysRevD.89.064007}, the requirement $\epsilon_0=0$ 
represents a sufficient condition to guarantee that Eq.~\eqref{JPmetric} 
satisfies asymptotic flatness at spatial infinity. Moreover, limits on the 
coefficients $\epsilon_{1,2}$ can be obtained from weak-field tests of 
gravity within the parametrized post-Newtonian framework, performed 
in the Solar System \cite{lrr-2006-3}. Such bounds translate into 
$\vert\epsilon_1\vert\lesssim10^{-5}$ and 
$\vert\epsilon_2\vert\lesssim4.6 \times10^{-4}$ \cite{PhysRevD.83.124015}. 

In this work, we assume $\epsilon_3$ as the only nonvanishing parameter 
of $h(r,\theta)$. Such a coefficient is currently unconstrained by observations 
and reflects changes of the Kerr metric at the order $\sim (M/r)^3$. We also 
focus on the orbital motion of massive test particles on equatorial circular 
geodesics, for which $\theta=\pi/2$. This condition further reduces 
Eq.~\eqref{JPh} to 
\begin{equation}\label{hreq}
h=\epsilon_3\frac{M^3r}{\Sigma^2}=\epsilon_3\frac{M^3}{r^3}\ .
\end{equation}

\begin{figure}[htbp]
\centering
\includegraphics[width=7.5cm]{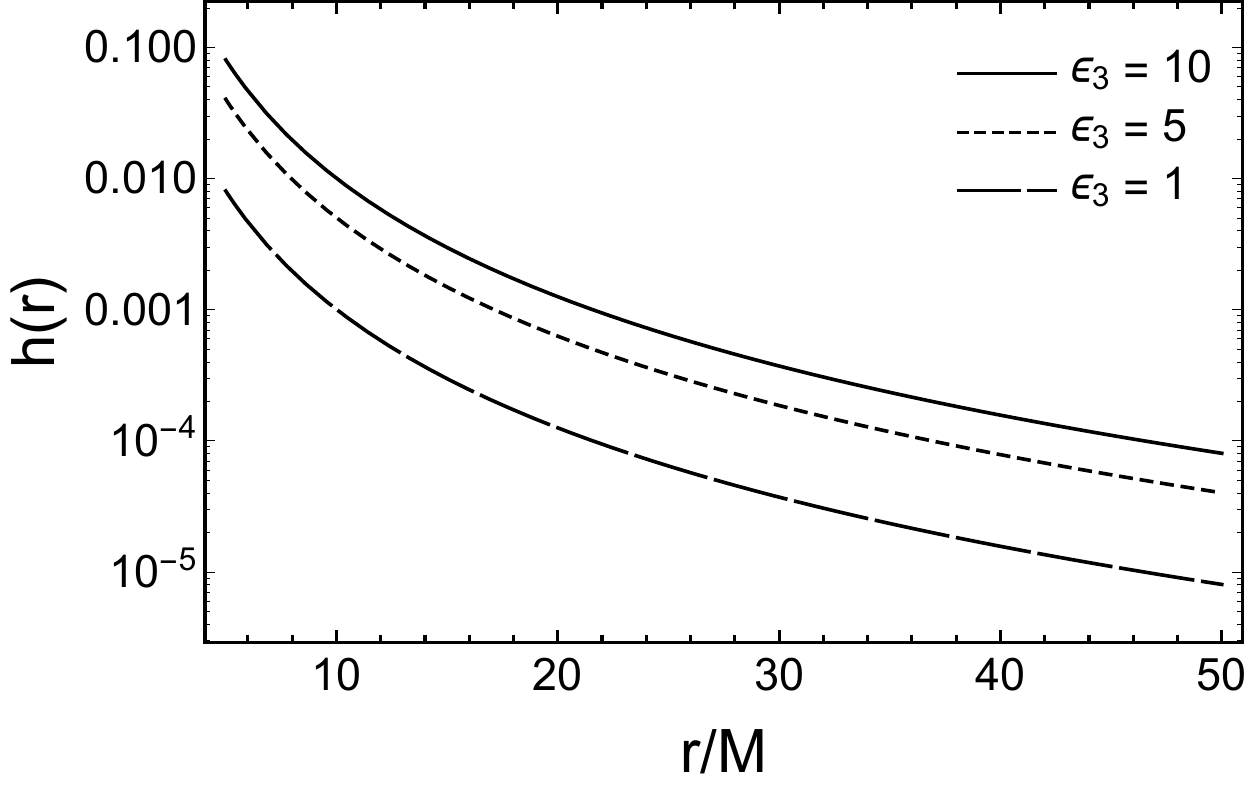}
\caption{We show the behavior of $h(r)$ of Eq.~\eqref{hreq} as function of the 
orbital distance normalised to the BH mass, for three values of the 
parameter $\epsilon_3=(1,5,10)$.}\label{fig1}
\end{figure}

Finally, we consider the strong-field effects identified by $h$, as small 
perturbations of the Kerr geometry, i.e. $\epsilon_3M^3/r^3\ll1$. We develop 
our framework at the linear approximation, neglecting $\mathcal{O}(h^2)$ 
corrections. The typical values $\epsilon_3$ considered in literature so far to 
analyze possible signatures of the JP metric are of the order $\sim\mathcal{O}(10)$
\cite{PhysRevD.83.124015,PhysRevD.87.124017,Bambi20115,PhysRevD.87.124010,PhysRevD.89.064007,1475-7516-2012-09-014}.
As shown in Fig.~\ref{fig1}, for such values, the condition $h\ll1$ is satisfied 
whenever the test particle orbits around the black hole at distances greater 
than $r\sim 6M$. This requirement is consistent with the study of astrophysical 
systems composed of a supermassive black hole and a solar-type star or a white 
dwarf, which are the primary target of our analysis \cite{Evans:2015fha}.
It is worth remarking that the parameter $\epsilon_3$ is 
expected to modify the BH quadrupole moment. This could affect 
the properties of geodesics around the central objects, varying the nodal 
precession frequency \cite{Pappas:2012nt,Pappas:2015mba,Gondek-Rosinska:2014aaa} 
or leading to vertical instabilities in the orbits \cite{Bambi:2011jq,Gair:2007kr}.\\

The JP metric Eq.~\eqref{JPmetric} is characterized by two killing 
vectors $k^\mu$ and $m^\mu$ associated to the space-time invariance with 
respect to time shifts and rotations along the polar angle $\phi$. The orbital 
motion of a test particle with 4-velocity $u^{\mu}=dx^\mu/d\tau=(u^t,0,0,u^\phi)$, $\tau$ being 
the proper time, is then featured by two conserved quantities related to $k^\mu$ and $m^\mu$,
\begin{align}
u^{\mu}k_{\mu}&=u^{t}g_{tt}+u^{\phi}g_{t\phi}=-{\cal E}\label{equt}\ ,\\ 
u^{\mu}m_{\mu}&=u^{\phi}g_{\phi\phi}+u^{t}g_{t\phi}=L\label{equphi} \ .
\end{align}
which can be identified with the energy at infinity and the angular 
momentum per unit mass of the test particle, respectively. An analytic 
expressions for ${\cal E}$ and $L$ may be derived solving the system 
of equations $V_\tn{eff}=0$ and $V'_\tn{eff}=0$,where 
\begin{equation}\label{effective}
V_\tn{eff}(r)=\frac{1}{g_{rr}}\left[\frac{g_{\phi\phi}{\cal E}^2+2g_{t\phi}{\cal E}L+g_{tt}L^2}{g_{t\phi}^2-g_{tt}g_{\phi\phi}}-1\right]
\end{equation}
is an effective potential governing the geodesic motion of a stationary 
and axisymmetric space-time \cite{Misner:1974qy}.  
From Eq.~\eqref{effective} and its derivative, we obtain at the linear 
order in $h$
\begin{align}
{\cal E}=\frac{1}{N}\bigg(1&-\frac{2M}{r}+a\omega_\tn{k}\bigg)-\frac{h}{N^3}
\left[\frac{3}{4}\frac{a^2}{r^2}+\frac{1}{4}\right.\nonumber\\
&\left.-ar^2\omega_\tn{k}^3-\frac{\omega_\tn{k}^2}{2}(r^2+6a^2)+\frac{3}{2}\frac{a^3}{r^2}\omega_\tn{k}
\right]\ ,
\end{align}
and
\begin{align}
L=\frac{r^2\omega_\tn{k}}{N}&\bigg(1+\frac{a^2}{r^2}-2a\omega_\tn{k}\bigg)-\frac{h}{N^3}
\bigg[r^2(a^2+3r^2)\omega^3_\tn{K}\nonumber\\
&+\left(\frac{3a^4}{2r^2}-a^2-3r^2\right)\omega_\tn{k}
+\frac{3}{4\omega_\tn{k}}\left(1+\frac{a^2}{r^2}\right)\nonumber\\
&-\left(3a^3+\frac{9ar^2}{2}\right)\omega^2_\tn{K}
+\frac{9a}{4r^2}(r^2+a^2)\bigg]\ ,
\end{align}
where $\omega_\tn{k}=(M/r^3)^{1/2}$ is the {\it Keplerian} orbital 
frequency, and we have defined 
$N=(1+2a\omega_\tn{k}-3r^2\omega^2_\tn{k})^{1/2}.$
Replacing the former expressions into Eqs.~\eqref{equt} and \eqref{equphi} 
leads to the 4-velocity components $u^t$ and $u^{\phi}$:
\begin{align}
u^t=&\frac{1+a\omega_\tn{k}}{N}-\frac{h}{N^34r^2}\left[3a^2+5r^2+6a(a^2+r^2)\omega_\tn{k}\nonumber\right.\\
&\left.-4r^2(4a^2+3r^2)\omega^2_\tn{K}+12ar^4\omega_\tn{k}^3\right]\ ,\\
u^\phi=&\frac{\omega_\tn{k}}{N}-\frac{h}{4r^2\omega_\tn{k}N^3}\left[6a^2\omega_\tn{k}^2
+a\omega_\tn{k}(9-16r^2\omega_\tn{k}^2)\right.\nonumber\\
&\left.+3(1-2r^2\omega_\tn{k}^2)^2\right]\ .
\end{align}


\section{Tidal field}\label{sec:tidalfield}

In GR, the effects of the stationary gravitational field are described by tidal forces 
acting on test masses. Single geodesics cannot detect gravity, and at least a pair 
of them is needed. In the presence of a mass $M$, the space-time is equipped with a 
metric $g_{\mu\nu}$, and a test body with 4-velocity $u^{\mu}$ will follow time-like 
geodesics of this metric. If we assume a second test particle, the position of which with 
respect to the first one is defined by the displacement vector $\delta x^\mu$, we 
can study the relative motion between them using a quasi-inertial Fermi coordinate 
system \cite{2006CQGra..23.4021C}. For this purpose, let us consider an orthonormal 
tetrad\footnote{Hereafter indices within round brackets will refer to tetrad components.} 
$\lambda^{\mu}{_{(\alpha)}}$ attached to $m$ (which stays forever at the origin of this 
coordinate frame), parallel transported along its geodesic parametrized by the proper 
time $\tau$. In Fermi coordinates, the metric at the second order in the displacement 
vector $y^{(i)}=\lambda^{(i)}{_{\mu}}\delta x^{\mu}$ will be given by
\begin{align}
g_{00}=&-1-R_{(0)(i)(0)(j)}y^{(i)}y^{(j)}+\ldots\ ,\\
g_{0i}=&-\frac{2}{3}R_{(0)(j)(i)(k)}y^{(j)}y^{(k)}+\ldots\ ,\\
g_{ij}=&\delta_{ij}-\frac{1}{3}R_{(i)(k)(j)(l)}y^{(k)}y^{(l)}+\ldots\ ,
\end{align}
where $R_{(\mu)(\nu)(\rho)(\sigma)}$ is the projection of the Riemann curvature 
tensor onto the orthonormal tetrad frame
\begin{equation}\label{RiemanProj}
R_{(\mu)(\nu)(\rho)(\sigma)}=R_{\alpha\beta\gamma\delta}\lambda^{\alpha}{_{(\mu)}}\lambda^{\beta}{_{(\nu)}}
\lambda^{\gamma}{_{(\rho)}}\lambda^{\delta}{_{(\sigma)}}\ .
\end{equation}
From this equation, we can define the {\it gravito-electric} 
and {\it gravito-magnetic} tidal tensors:
\begin{align}
{\cal E}_{(i)(j)}=&R_{(0)(i)(0)(j)}\label{gravitoelec} \ ,\\
 {\cal H}_{(i)(j)}=&-\frac{1}{2}\epsilon_{ikl}R^{(k)(l)}{_{(0)(j)}}\label{gravitomagn}\ ,
\end{align} 
with $\epsilon_{ijk}$ being the Levi-Civita symbol. The electric component 
${\cal E}_{(i)(j)}$ describes tidal deviations of nearby geodesics, while the 
magnetic term ${\cal H}_{(i)(j)}$ is directly related to frame-dragging effects 
of test gyroscopes. Both tensors are symmetric and trace free. This setup is 
physically equivalent, in case of negligible Fermi velocity, to solving the geodesic 
deviation equation, which can be written in the tetrad frame 
$\lambda^{\mu}{_{(\alpha)}}$ as \cite{1956AcPP...15..389P}
\begin{equation}
\frac{d^2y_{(i)}}{d\tau^2}+{\cal E}_{(i)(j)}y^{(j)}=0\ .
\end{equation}
In this section, we shall derive the expressions for Eqs.~\eqref{gravitoelec} and \eqref{gravitomagn} 
in the JP metric, following the approach adopted from \cite{2006CQGra..23.4021C}.
To determine the parallel transported vectors $\lambda^{\mu}{_{(\alpha)}}$, 
we first consider the tetrad $e^{\mu}{_{(\alpha)}}$ associated to a static observer 
in the space-time \eqref{JPmetric}, such that the line element takes the form
\begin{equation}
ds^2=\eta_{\alpha\beta}d\xi^{(\alpha)}d\xi^{(\beta)}\ ,
\end{equation}
where $\eta_{\alpha\beta}=\tn{diag}(-1,1,1,1)$ is the flat-space metric tensor 
and $d\xi^{(\alpha)}=e^{(\alpha)}{_{\mu}}dx^{\mu}$. We immediately note 
from Eq.~\eqref{JPmetric} that for  $e^{(1)}{_{\mu}}$ and $e^{(2)}{_{\mu}}$ 
the basis vectors reduce to
\begin{align}
e^{(1)}{_{\mu}}=&\left(0,\frac{r}{\Delta^{1/2}}\left[1+\frac{r^2fh}{2\Delta}\right],0,0\right)\ ,\\
e^{(2)}{_{\mu}}=&(0,0,r,0)\ .
\end{align}
The other two components can be derived from the orthogonality condition 
$\eta_{\alpha\beta}e^{(\alpha)}{_{\mu}}e^{(\alpha)}{_{\mu}}=g_{\mu\nu}$,
from which we find:
\begin{align}
e^{(0)}{_{\mu}}=&\left(f[1+h/2],0,0,\frac{2aM}{rf}[1+h/2]\right)\ ,\\
e^{(3)}{_{\mu}}=&\left(0,0,0,\frac{\sqrt{\Delta}}{f}+\frac{a^2h}{2f\Delta^{1/2}}\right)\ ,
\end{align}
where $f=\sqrt{1-2M/r}$.

The parallel transported tetrad can be now obtained applying to $\lambda^{\mu}{_{(\alpha)}}$ 
a Lorentz boost along the $3$-direction, such that the time component of 
the new reference frame coincides with the test particle 4-velocity $u^{\mu}$.
The new basis vectors on the worldline read 
\begin{align}
\lambda^{\mu}{_{(0)}}=&\gamma\left[e^{\mu}{_{(0)}}+\beta e^{\mu}{_{(3)}}\right]\quad\ ,
\quad\lambda^{\mu}{_{(1)}}=e^{\mu}{_{(1)}}\ ,\\
\phantom{a}\nonumber\\
\lambda^{\mu}{_{(2)}}=&e^{\mu}{_{(2)}}\quad\ ,\quad\lambda^{\mu}{_{(3)}}
=\gamma\left[e^{\mu}{_{(0)}}+\beta e^{\mu}{_{(3)}}\right]\ ,
\end{align}
where $\beta$ and $\gamma$ are the boost velocity and the corresponding 
Lorentz factor. They can be easily obtained from the condition $\lambda^{\mu}{_{(0)}}=u^{\mu}$, 
from which we find
\begin{align}
\gamma=&\frac{1}{Nf}[1+\omega_\tn{k}(a-2r^2\omega_\tn{k})]-\frac{h}{4r^2N^3f}\left[6a^3\omega_\tn{k}
\right.\nonumber\\
&+a^2(3-8r^2\omega_\tn{k}^2)+2ar^2\omega_\tn{k}(3-5r^2\omega_\tn{k}^2)\nonumber\\
&\left.+3r^2(1-2r^2\omega_\tn{k}^2)^2\right]\label{betafactor}\ ,\\
\beta=&\frac{\omega_\tn{k}\Delta^{1/2}}{1+\omega_\tn{k}(a-2r^2\omega_\tn{k})}-\frac{hf^2}{4\omega_\tn{k}}
\frac{\Delta^{-1/2}}{(r+ar\omega_\tn{k}-2M)^2}\nonumber\\
&\times\left[a^2(3-8r^2\omega_\tn{k}^2)+2ar^2\omega_\tn{k}(3-5r^2\omega_\tn{k}^2)\right.\nonumber\\
&\left.+6a^3\omega_\tn{k}+3r^2f^4\right]\label{gammafactor}\ .
\end{align}
The time component of the tetrad $\lambda^\mu{_{(0)}}$ is automatically parallel 
transported along the particle wordline, as the vector $\lambda^\mu{_{(2)}}$. 
However, the two spatial vectors $\lambda^{\mu}{_{(1)}}$ and $\lambda^{\mu}{_{(3)}}$, 
which in spherical polar coordinates are along the radial and tangential directions 
with respect to the orbit, must to be rotated to be parallel propagated. Therefore, 
we introduce two new vectors $\tilde{\lambda}^{\mu}{_{(1)}}, \tilde{\lambda}^{\mu}{_{(3)}}$, 
defined as:
\begin{align}
\tilde{\lambda}^{\mu}{_{(1)}}&=\lambda^{\mu}{_{(1)}}\cos\xi-\lambda^{\mu}{_{(3)}}\sin\xi\ ,\\
\tilde{\lambda}^{\mu}{_{(3)}}&=\lambda^{\mu}{_{(1)}}\sin\xi+\lambda^{\mu}{_{(3)}}\cos\xi\ .
\end{align}
Requiring that $\tilde{\lambda}^{\mu}{_{(1)}}$ and $\tilde{\lambda}^{\mu}{_{(3)}}$ 
satisfy the parallel transport 
equation along the wordline with tangent vector $\lambda^{\mu}{_{(0)}}$
\begin{equation}
\lambda^{\nu}{_{(0)}}(\nabla_{\nu}\tilde{\lambda}^{\mu}{_{(1)}})=\lambda^{\nu}{_{(0)}}(\nabla_{\nu}\tilde{\lambda}^{\mu}{_{(3)}})=0\ ,
\end{equation}
yields for the $\xi$:
\begin{equation}\label{lagangle}
\xi=\omega_\tn{k}\tau\left(1-\frac{3r-4M}{4M}h\right)\ ,
\end{equation}
having fixed the integration constant such that $\xi(\tau=0)=0$.
The full expression for the basis vectors $(\lambda^\mu{_{(0)}},\tilde{\lambda}^\mu{_{(1)}},\lambda^\mu{_{(2)}},\tilde{\lambda}^\mu{_{(3)}})$
at the linear order in $h$ is given in the Appendix~\ref{AppBasis}.

Having computed the parallel transported tetrad, we can now project the 
$R_{\alpha\beta\gamma\delta}$ to derive the gravito-electric and gravito-magnetic 
tensors \eqref{gravitoelec}-\eqref{gravitomagn}. For the sake of clarity, we split 
each component as a sum of two pieces: one related to the pure Kerr geometry 
and one corresponding to the corrections induced by the parameter $h$ in the 
JP metric,
\begin{equation}
{\cal E}_{(i)(j)}=\bar{{\cal E}}_{(i)(j)}+h\ \delta{\cal E}_{(i)(j)}\ ,
\end{equation}
where
\begin{align}
\bar{{\cal E}}_{(1)(1)}=&\omega_\tn{k}^2\left(1-\frac{3}{r^2}\frac{\Delta}{N^2}\cos^2\xi\right)\label{E110}\ ,\\
\bar{{\cal E}}_{(2)(2)}=&\frac{\omega_\tn{k}^2}{N^2}\left(1+3\frac{a^2}{r^2}-4a\omega_\tn{k}\right)\ ,\\
\bar{{\cal E}}_{(3)(3)}=&\omega_\tn{k}^2\left(1-\frac{3}{r^2}\frac{\Delta}{N^2}\sin^2\xi\right)\ ,\\
\bar{{\cal E}}_{(1)(3)}=&-\frac{3}{2}\frac{\Delta}{N^2r^2}\omega_\tn{k}^2\sin2\xi\ ,\label{E130}
\end{align}
with $\bar{{\cal E}}_{22}=-(\bar{{\cal E}}_{11}+\bar{{\cal E}}_{33})$ and 
$\xi$ given by Eq.~\eqref{lagangle}. The changes to the electric tidal tensor induced by the strong 
field deviations read
\begin{widetext}
\begin{align}
\delta{\cal E}_{(1)(1)}=&\frac{4\omega_\tn{k}^2r^2-3}{2r^2}\sin^2\xi
+\frac{1}{2r^4N^4}\left[3(5a^2+4r^2)+(51a^3+39ar^2)\omega_\tn{k}+(42a^4-48a^2r^2-80r^4)\omega_\tn{k}^2\right.\nonumber\\
&\left.-2(66a^3r^2+85a r^4)\omega_\tn{k}^3+(40a^2r^4+183 r^6)\omega_\tn{k}^4+192a r^6\omega^5_\tn{K}-144r^8\omega_\tn{k}^6\right]\cos^2\xi\ ,\label{dE11}\\
\delta{\cal E}_{(2)(2)}=&\frac{1}{N^4r^2}\bigg[-\frac{3}{2}\left(3+\frac{5a^2}{r^2}\right)-\frac{3}{2}a\left(9+\frac{17a^2}{r^2}\right)\omega_\tn{k}
+\left(30a^2-\frac{21a^4}{r^2}+29r^2\right)\omega^2_\tn{K}+(66a^3+59a r^2)\omega_\tn{k}^3\nonumber\\
&-(28a^2r^2+66r^4)\omega_\tn{k}^4-72ar^4\omega_\tn{k}^5+54\omega_\tn{k}^6r^6\bigg]\ ,\\
\delta{\cal E}_{(3)(3)}=&\frac{4\omega_\tn{k}^2r^2-3}{2r^2}\cos^2\xi
+\frac{1}{2r^4N^4}\left[3(5a^2+4r^2)+(51a^3+39ar^2)\omega_\tn{k}+(42a^4-48a^2r^2-80r^4)\omega_\tn{k}^2\right.\nonumber\\
&\left.-2(66a^3r^2+85a r^4)\omega_\tn{k}^3+(40a^2r^4+183 r^6)\omega_\tn{k}^4+192a r^6\omega^5_\tn{K}-144r^8\omega_\tn{k}^6\right]\sin^2\xi\ ,\\
\delta{\cal E}_{(1)(3)}=&\frac{1}{N^4r^4}\bigg[\frac{15}{4}(a^2+r^2)+
\frac{51}{4}a(r^2+a^2)\omega_\tn{k}+\frac{3}{2}(7a^4-6a^2r^2-17r^4)\omega_\tn{k}^2-\frac{3}{2}(22a^3r^2+37ar^4)\omega_\tn{k}^3\nonumber\\
&+\left(6a^2r^4+\frac{117}{2}r^6\right)\omega_\tn{k}^4+60ar^6\omega_\tn{k}^5-45r^8\omega_\tn{k}^6\bigg]
\sin2\xi\ .\label{dE13}
\end{align}
\end{widetext}

Similarly, the nonvanishing components of the magnetic term \eqref{gravitomagn} 
are :
\begin{align}
\bar{{\cal H}}_{(1)(2)}=&-3\frac{\omega_\tn{k}^2}{r^2}(\ok r^2-a)\frac{\Delta^{1/2}}{N^2}\cos\xi\ ,\\
\bar{{\cal H}}_{(2)(3)}=&-3\frac{\omega_\tn{k}^2}{r^2}(\ok r^2-a)\frac{\Delta^{1/2}}{N^2}\sin\xi\ ,
\end{align}
and finally
\begin{widetext}
\begin{align}
\delta{\cal H}_{(1)(2)}=&-\frac{\Delta^{-1/2}}{r^2N^4}\bigg[\left(33a^2+\frac{18a^4}{r^2}
+11r^2\right)\frac{\ok}{4}+\bigg(\frac{27}{2}a^3+\frac{9a^5}{r^2}+2ar^2\bigg)\ok^2-\frac{\ok^3}{4}\left(72a^4+163a^2r^2+55r^4\right)\nonumber\\
&-\left(18a^3r^2-\frac{ar^4}{2}\right)\ok^4
+(47a^2r^4+21r^6)\ok^5-9ar^6\ok^6-9r^8\ok^7\bigg]\cos\xi\ ,\label{dH12}\\
\delta{\cal H}_{(2)(3)}=&-\frac{\Delta^{-1/2}}{r^2N^4}\bigg[\left(45a^2+\frac{18a^4}{r^2}+27r^2\right)\frac{\ok}{4}
+\bigg(\frac{51a^3}{2}+\frac{9a^5}{r^2}+15ar^2\bigg)\ok^2-\left(24a^4+195a^2r^2+171r^4\right)\frac{\ok^3}{4}\nonumber\\
&-\left(54a^3r^2+\frac{117ar^4}{2}\right)\ok^4+(54a^2r^4+90r^6)\ok^5+57ar^6\ok^6-63r^8\ok^7\bigg]\sin\xi\ .\label{dH23}
\end{align}
\end{widetext}


\subsection{Relevance of strong-gravity corrections}\label{sec:results}

\begin{figure}[htbp]
\centering
\includegraphics[width=8.5cm]{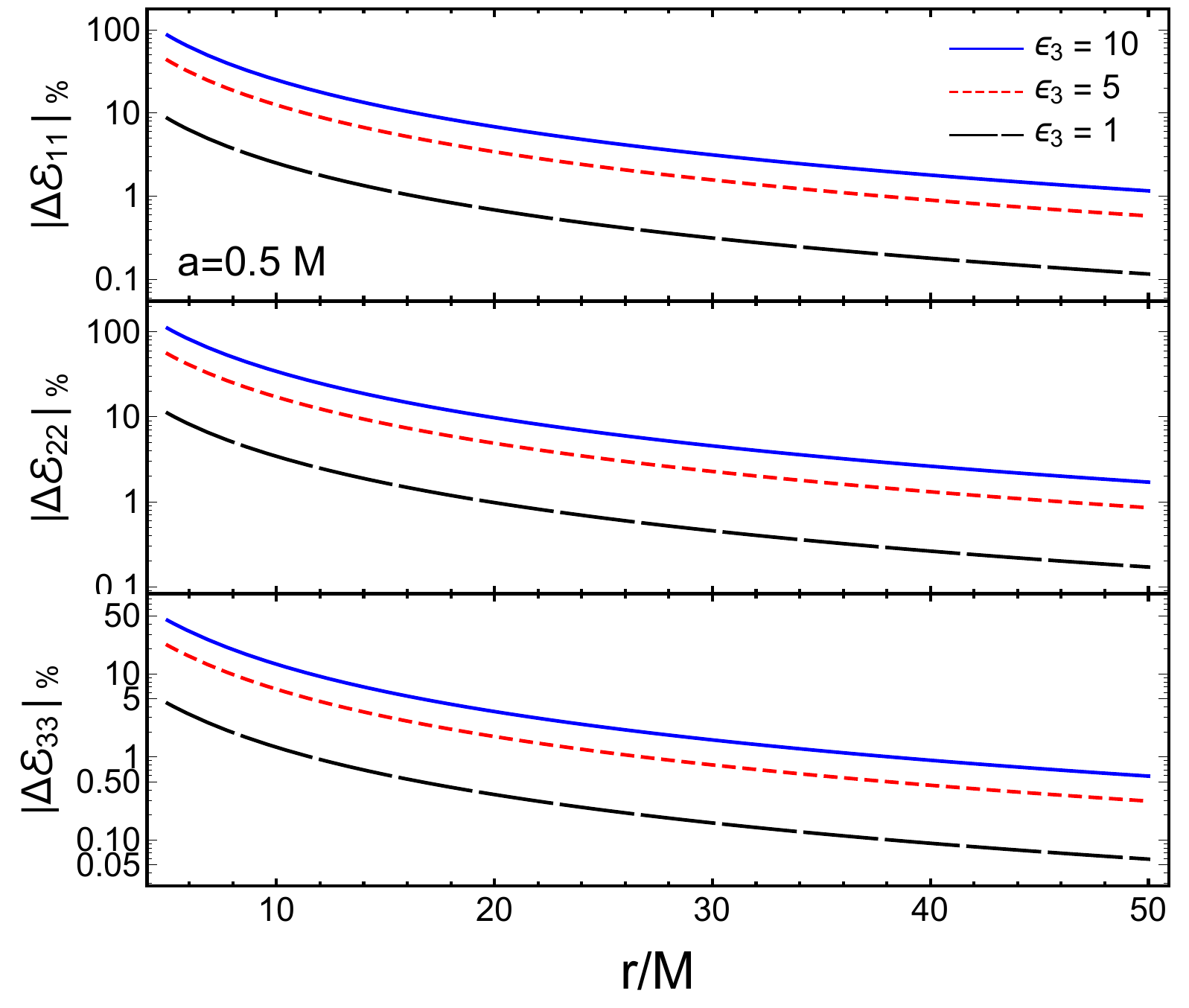}
\caption{In this figure, we plot the absolute percentage values of 
$\Delta{\cal E}_{ij}$ given by Eq.~\eqref{deltaEH} as a function of the orbital 
distance, for three values of the strong-gravity parameter $\epsilon_3=(1,5,10)$, 
and BH spin $a=0.5M$.}\label{fig2}
\end{figure}

\begin{figure}[htbp]
\centering
\includegraphics[width=8cm]{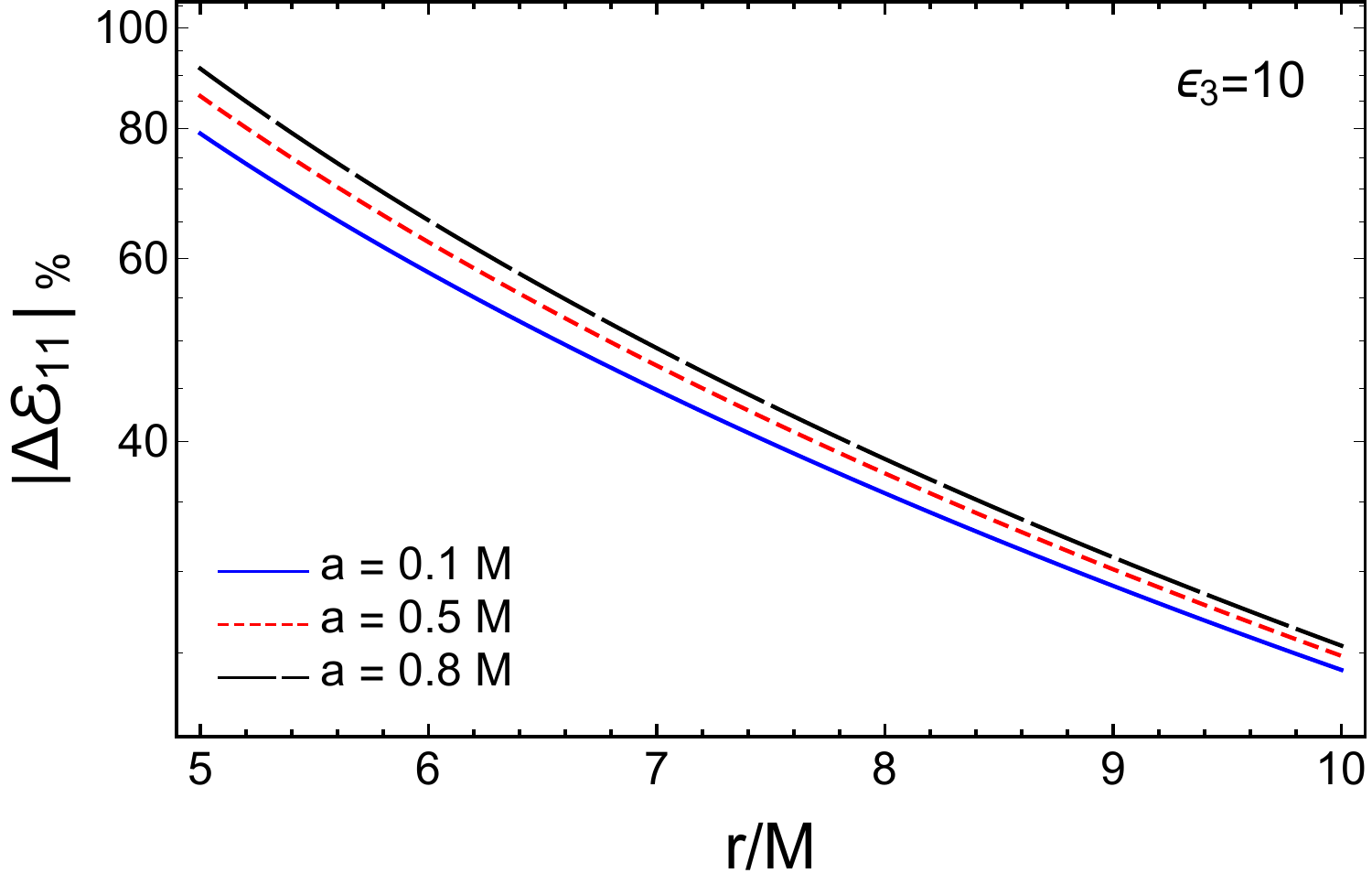}
\caption{Same as Fig.~\ref{fig2} but only for $\Delta{\cal E}_{11}$, with 
$\epsilon_3=10$ and different values of the BH spin parameter 
$a/M=(0.1,0.5,0.8)$.}
\label{fig3}
\end{figure}

\begin{figure}[htbp]
\centering
\includegraphics[width=8cm]{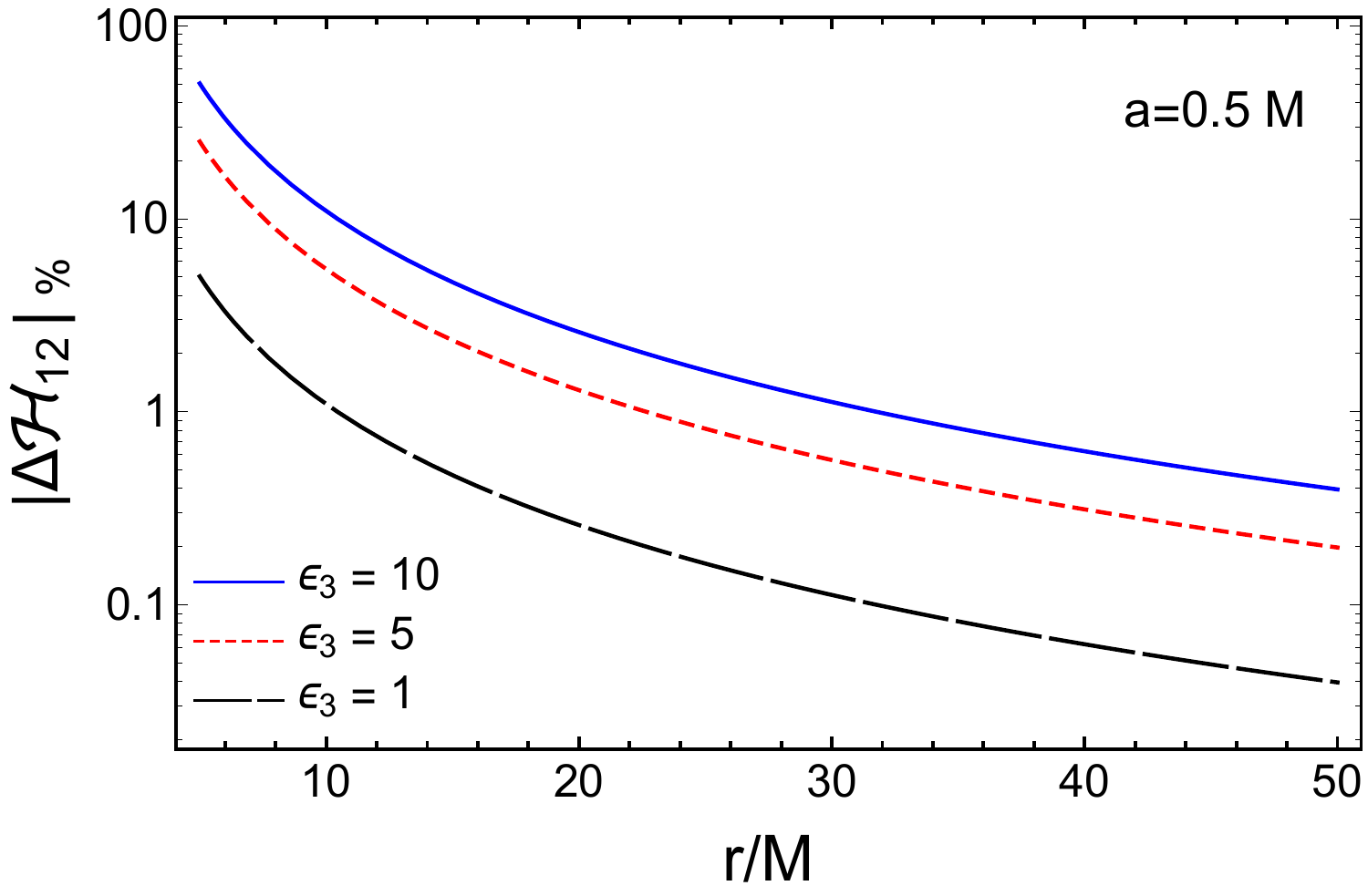}
\caption{Same as Fig.~\ref{fig2} but for the magnetic component 
$\Delta{\cal H}_{12}$.}
\label{fig4}
\end{figure}

Given the explicit form of the gravito-magnetic and electric tidal tensors 
Eqs.~\eqref{E110}-\eqref{dH23}, we need to estimate the relevance of the 
non-Kerr components as function of the BH angular momentum and the 
deformation parameter $\epsilon_3$. To this aim, we define the two quantities 
\begin{equation}
\Delta{\cal E}_{ij}=h\frac{\delta{\cal E}_{(i)(j)}}{\bar{{\cal E}}_{(i)(j)}}\qquad\tn{and}\qquad
\Delta{\cal H}_{ij}=h\frac{\delta{\cal H}_{(i)(j)}}{\bar{{\cal H}}_{(i)(j)}}\ ,\label{deltaEH}
\end{equation}
which represent the fractional change with respect to ${\cal E}$ and ${\cal H}$ 
computed in the standard GR scenario. In the following, starting from an initial  
configuration with $\xi(\tau_0)=0$, we consider snapshots at different orbital 
distances with the same phase $\xi(\tau)=\xi(\tau_0)$. Even though a more accurate 
analysis will be developed in the next section through a numerical approach, this 
assumption will provide, as first hint, an order of magnitude estimate of the effects 
we are going to study. We note that in this case ${\cal E}_{(1)(3)}={\cal H}_{(2)(3)}=0$. 
Our results can be summarized in Figs. \ref{fig2}-\ref{fig4}.

In the three panels of Fig.~\ref{fig2}, we show the absolute value of $\Delta{\cal E}_{ij}$ 
as a function of the orbital distance, for $\epsilon_3=(1,5,10)$ and BH spin parameter 
$a/M=0.5$. As expected, the contribution of strong-gravity terms grows as $r$ 
decreases and can be of the order $\sim 10\%$ for $r>10M$. 
For $r<6M$ and $\epsilon_3>5$, the relative difference is always larger than $50\%$; 
terms of second order $\mathcal{O}(h^2)$ and proportional to higher coefficients as $\epsilon_{4}$ 
start to be relevant and cannot be neglected. For $\epsilon_3>0$ ($<0$ respectively), 
all the components of $\Delta{\cal E}_{ij}$ are smaller (higher) than zero, and therefore 
the strong-gravity corrections reduce (increase) the neat effect of tidal deviations induced 
by the gravito-electric tensor\footnote{This feature could be qualitatively expected since 
for $\epsilon_3>0$ ($\epsilon_3<0$) the modified BH is more prolate (oblate) than the 
Kerr one \cite{Bambi20115}.}. 

In Fig.~\ref{fig3}, we draw $\Delta{\cal E}_{11}$ for $\epsilon_3=10$ and different values 
of $a/M=(0.1,0.5,0.8)$. The plot shows that unless the binary systems gets very close at 
orbital distances $r\ll 10M$, the effects of non-Kerr deviations seem to be insensitive to 
the BH spin. This feature does not change for the other components of ${\cal E}_{(i)(j)}$.

\begin{figure}[htbp]
\centering
\includegraphics[width=8cm]{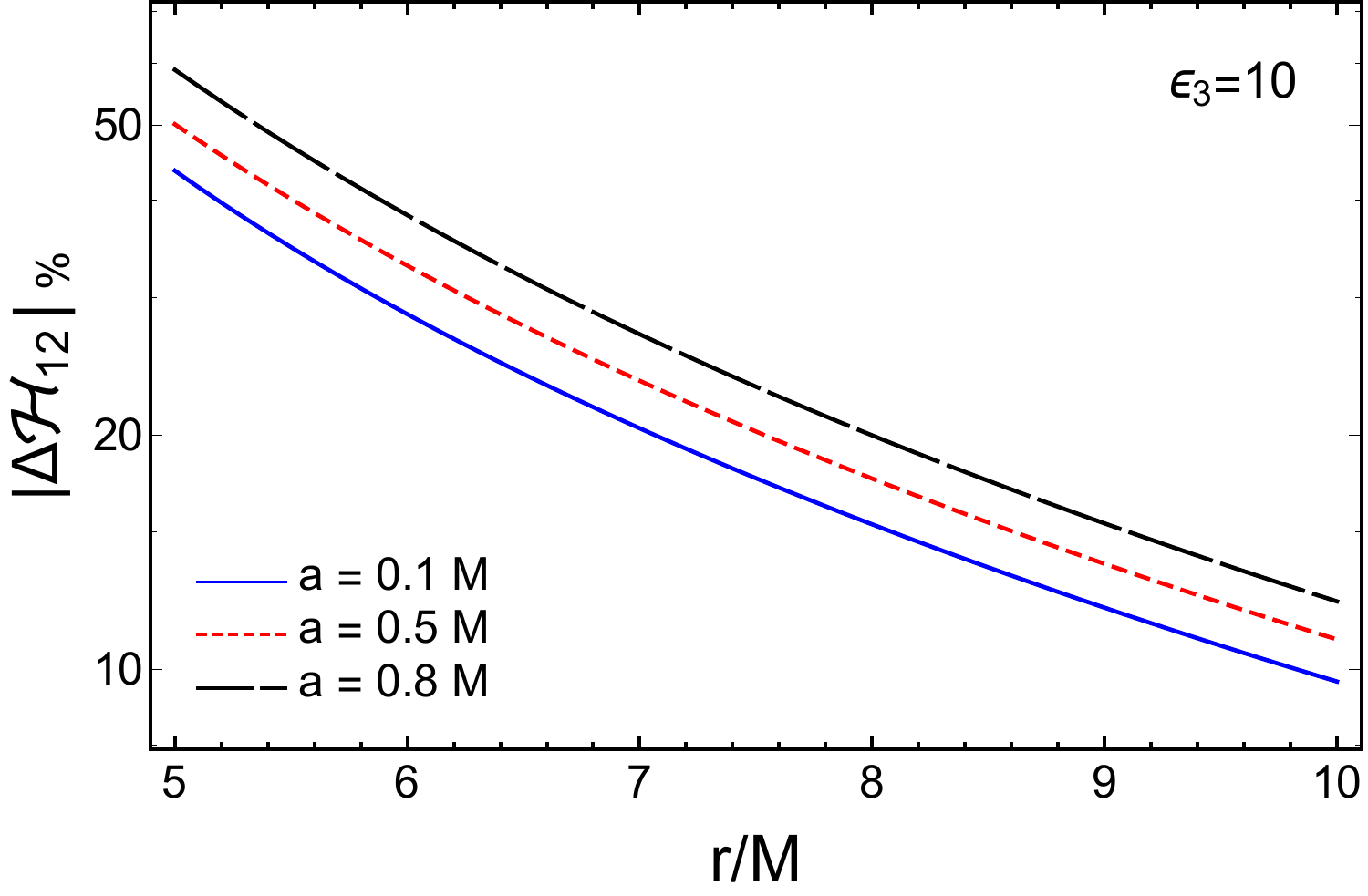}
\caption{Same as Fig.~\ref{fig3} but for the magnetic component 
$\Delta{\cal H}_{12}$.}
\label{fig5}
\end{figure}

Finally, we note that the picture described above also applies to the gravito-magnetic 
tidal tensor. We show the behavior of $\Delta{\cal H}_{12}$ in Figs.~\ref{fig4} and 
\ref{fig5} for the same set of parameters previously considered.

\section{Black hole-white dwarf binary evolution}\label{sec:numerical}
Hereafter, we will analyze the effects of strong-field corrections derived in the previous 
section, using our results together with the affine model, which represents a semianalytic 
approach to describing stellar deformations induced by an external tidal field. We shall first 
provide a brief summary of the main ingredients of the model, referring the reader to Ref. 
\cite{Maselli:2012zq} (and reference therein) for a more comprehensive description of this 
framework. Then, we will present the numerical results obtained for different configurations 
of prototype BH-WD binaries.

\subsection{Model}
The main assumption of the affine approach is that the spherical star is deformed by the 
tidal field into an ellipsoid, preserving this shape during the orbital motion. More specifically, 
it is warped in an S-type Riemann ellipsoid, for which the spin and vorticity are parallel, and 
their ratio constant \cite{trove.nla.gov.au/work/19653231}. The equations for the stellar 
deformations are written in the {\it principal frame}, which is comoving with the star, and such 
that the axes are adapted to the principal axes of the ellipsoid \footnote{In the following, "1" 
denotes the direction along the axis parallel to the orbital separation, "2" identifies the axis 
orthogonal to the orbital plane, and "3" defines the other axis in the orbital plane.} 
$a_{i=1,2,3}$.  Under this assumption, the infinite number of degrees of freedom of the internal 
fluid, is reduced to a set of five variables $(a_1,a_2,a_3,\psi,\lambda)$, where $\psi$ and 
$\lambda$ are the two angles:
\begin{equation}\label{Affienangles}
\frac{d\psi}{d\tau}=\Omega\qquad\ ,\qquad \frac{d\lambda}{d\tau}=\Lambda\ .
\end{equation}
In the previous expressions, $\Omega$ is the WD angular velocity measured in the 
tetrad coordinate system (which is parallel transported), and $\Lambda$ describes the 
internal fluid motion in the principal frame.

The equations of motion for the star can be derived from the 
Lagrangian
\begin{equation}\label{Lagrangian}
{\cal L}={\cal L}_\tn{B}+{\cal L}_\tn{T} \ ,
\end{equation}
where the subscripts T and B refer to {\it tidal} and {\it body}. The first term reads
\begin{equation}
{\cal L}_\tn{T}=-\frac{1}{2}c_{ij}I_{ij}\ ,
\end{equation}
where $I_{ij}$ s the inertial tensor, which is written in the 
affine model as
\begin{equation}
I_{ij}=\hat{\cal M}\cdot \tn{diag}\left(\frac{a_{i}}{R_\tn{WD}}\right)^2
=\frac{4\pi}{3}\int_0^{R_\tn{WD}}\hat{\rho}(r) r^4 dr\ ,
\end{equation}
$\hat{\cal M}$ being the scalar quadrupole moment computed over the density 
profile of the star\footnote{The superscript hat denotes quantities computed for the 
spherical star.} and $R_\tn{WD}$ its radius at spherical equilibrium. 
$c_{ij}$ are the components of the gravito-electric tidal tensor in the principal 
frame, obtained by rotating ${\cal E}_{ij}$ of the angle $\psi$ [defined in Eq.~\eqref{Affienangles}] 
$c=T{\cal E}T^{T}$, where the matrix $T$ is given by
\begin{equation}
T=
\begin{pmatrix}
\cos\psi&0 &\sin\psi\\
0 &1 &0\\
-\sin\psi &0 &\cos\psi
\end{pmatrix}\ .
\end{equation}
Practically speaking, this rotation changes the angle $\xi$ into $l=\psi-\xi$. 
The latter describes the misalignment between the $a_1$ axis and the line between 
the two objects: when $l$ is negligible, the binary is said to be {\it synchronized}. 
For a star with zero viscosity, this angle is in general very small. However, as the orbit 
shrinks, the tidal bulge lags behind the tidal potentia, and then is subject to a torque 
which tries to spin it to follow the orbital motion, and the system desynchronizes 
\cite{Lai:1993rs,Dall'Osso:2012hm}. 

The body Lagrangian ${\cal L}_\tn{B}$ describes the star internal dynamics and 
contains three contributions coming from the kinetic, the internal energy of the fluid, and 
the self-gravity (see Ref. \cite{Maselli:2012zq} for a complete expression of these 
quantities in terms of the affine model variables).
By applying the Euler-Lagrange formalism to Eq.~\eqref{Lagrangian}, we obtain the 
equations of motion for the star
\begin{align}
\ddot{a}_1=&a_1(\Lambda^2+\Omega^2)-2a_2\Lambda\Omega+
\frac{1}{2}\frac{\hat{V}}{\hat{\cal M}}R_\tn{WD}^3a_1 A_1\nonumber\\
-&\frac{R^2_\tn{WD}}{{\cal M}}\frac{\hat{V}}{3a_1}-c_{11}a_1\label{eqaffine1}\ ,\\
\ddot{a}_3=&a_3(\Lambda^2+\Omega^2)-2a_1\Lambda\Omega+
\frac{1}{2}\frac{\hat{V}}{\hat{\cal M}}R_\tn{WD}^3a_3 A_3\nonumber\\
-&\frac{R^2_\tn{WD}}{\hat{\cal M}}\frac{\hat{V}}{3a_3}-c_{33}a_3\ ,\\
\ddot{a}_2=&\frac{1}{2}\frac{\hat{V}}{\hat{\cal M}}R_\tn{WD}^3a_2 A_2
-\frac{R^2_\tn{WD}}{\hat{\cal M}}\frac{\hat{V}}{3a_2}-c_{22}a_2\ ,\\
\dot{J}=&\frac{\hat{\cal M}}{R^2_\tn{WD}}c_{13}(a_3^2-a_1^2)\ ,\\
\dot{{\cal C}}=&0 \label{eqaffine5}\ ,
\end{align}
where a dot refers to differentiation with respect to the proper time $\tau$ and 
$\hat{V}$ is the star self-gravity at spherical equilibrium, given by
\begin{equation}\label{selfgrav}
\hat{V}=-\frac{G}{2}\int_\tn{spher} r\partial_r\Phi_\tn{Newt}dm\ ,
\end{equation}
with $\Phi_\tn{Newt}$ the Newtonian gravitational potential and $dm$ 
the WD mass element. We have also introduced the quantities
\begin{align}
A_{i}=&\int_0^{\infty}\frac{du}{(a_i^2+u)\sqrt{(a_1^2+u)(a_2^2+u)(a_3^2+u)}}\ ,\\
J=&\frac{\hat{\cal M}}{R^2_\tn{WD}}[(a_1^2+a_3^2)\Omega-2a_1a_3\Lambda]\ ,\\
{\cal C}=&\frac{\hat{\cal M}}{R^2_\tn{WD}}[(a_1^2+a_3^2)\Lambda-2a_1a_3\Omega]\ ,
\end{align}
$J$ being the star angular momentum and ${\cal C}$ the circulation of the fluid 
\cite{0004-637X-532-1-530}. In the absence of viscosity, as for the models we are going 
to study here, ${\cal C}$ is a constant of motion. We also consider 
irrotational configurations, for which ${\cal C}=0$. 

In this work, we investigate quasiequilibrium sequences of BH-WD 
binaries; this assumption reduces Eqs~\eqref{eqaffine1}-\eqref{eqaffine5} to 
a system of coupled algebraic equations,
\begin{equation}\label{eqaffineeq}
\ddot{a}_i=0 \quad\ ,\quad \psi=\xi\quad\ ,\quad\dot{\psi}=\Omega=\dot{\xi}\ ,
\end{equation}
which is solved through a Newton-Raphson method. 

\subsection{Numerical results}

We employ the affine model for a representative set of binary configurations. 
The WD equilibrium structure is built within a Newtonian framework using a polytropic 
equation of state $P=K\rho^\gamma$ with $\gamma =5/3$, where $P$ and $\rho$ are 
the pressure and mass-density profiles. We choose the central density 
and the constant $K$ such that the star has mass and radius $M_\tn{WD}=1 M_\odot$ 
and $R_\tn{WD}=7088$ km. 
Moreover, we consider rotating BHs with mass $M=10^4 M_\odot$, spins $a=(0.5,0.8)M$, 
and five values for the deformation parameter of the JP metric $\epsilon_3=(0,\pm 5,\pm10)$.
For each configuration, we solve the system of Eqs.~\eqref{eqaffineeq} placing the spherical 
star at the orbital separation $r\gg R_\tn{WD}$ from the BH. Then, we gradually reduce the 
distance until it reaches the critical point $r_\tn{tide}$ at which the WD fills its Roche lobe. 
The latter defines the region around the star in which a particle with mass $m\ll M_\tn{WD}$ 
is gravitationally bounded to the central object. At the Newtonian level, the Roche lobe 
can be identified by finding the maximum of the three-body potential (in the equatorial plane 
$x$-$y$)
\begin{equation}\label{3bodypot}
U(x,y)=-\frac{G m_1}{\vert\vec{x}-\vec{y}_1\vert}-\frac{G m_1}{\vert\vec{x}-\vec{y}_2\vert}
-\frac{G}{2}\frac{(m_1+m_2)}{\vert\vec{y}_1-\vec{y}_2\vert^3} x^2\ ,
\end{equation}
where $\vec{y}_{1/2}$ are the displacement vectors and in our case 
$m_1=M$, $m_2=M_\tn{WD}$. 
At each step of the simulation, we numerically compute Eq.~\eqref{3bodypot} and its maximum, 
defining $r_\tn{tide}$ as the orbital distance for which the WD axis $a_1$, elongated by the tidal forces, 
touches the Roche lobe.
 
Our results can be summarized as follows:
\begin{itemize}
\item 
In Fig.~\ref{fig6}, we show the relative difference between the axes $a_{1},a_2$ computed for 
$\epsilon_{3}=(\pm5,\pm10)$ and $\epsilon_3=0$, namely,
\begin{equation}\label{deltaa}
\Delta a_1=\frac{a_1\big\vert_{\epsilon_3=\pm5}}{a_1\big\vert_{\epsilon_3=0}}-1\quad\ ,\quad 
\Delta a_1=\frac{a_1\big\vert_{\epsilon_3=\pm10}}{a_1\big\vert_{\epsilon_3=0}}-1\ ,
\end{equation}
as function of the orbital distance normalized to the BH mass, for $a/M=0.5$. 
This quantity is evaluated  up to the radius\footnote{As noted in Sec.~\ref{sec:results}, 
values of the parameter $\epsilon_3$ greater than zero reduce the effect of the tidal 
field and then make the star fill its Roche lobe at smaller orbital distances 
than the pure GR case for which $\epsilon_3=0$. Vice versa, for $\epsilon_3<0$, 
the star touches the Roche lobe surface earlier.} 
$r_\tn{tide}\big\vert_{\epsilon_3=0}$ 
or $r_\tn{tide}\big\vert_{\epsilon_3=-5,-10}$, for positive and negative values of $\epsilon_3$, 
respectively.  
As the relative separation shrinks, the difference between 
the GR and the alternative scenario increases up to $\sim 3\%$ and $\sim 5\%$ for 
$\epsilon_3=\pm5$ and $\epsilon_3=\pm10$, respectively. The effect on the axis $a_2$ 
is less pronounced with discrepancies smaller than $1\%$; 
the same results apply to the axis $a_3$. 
\begin{figure}[htbp]
\centering
\includegraphics[width=8cm]{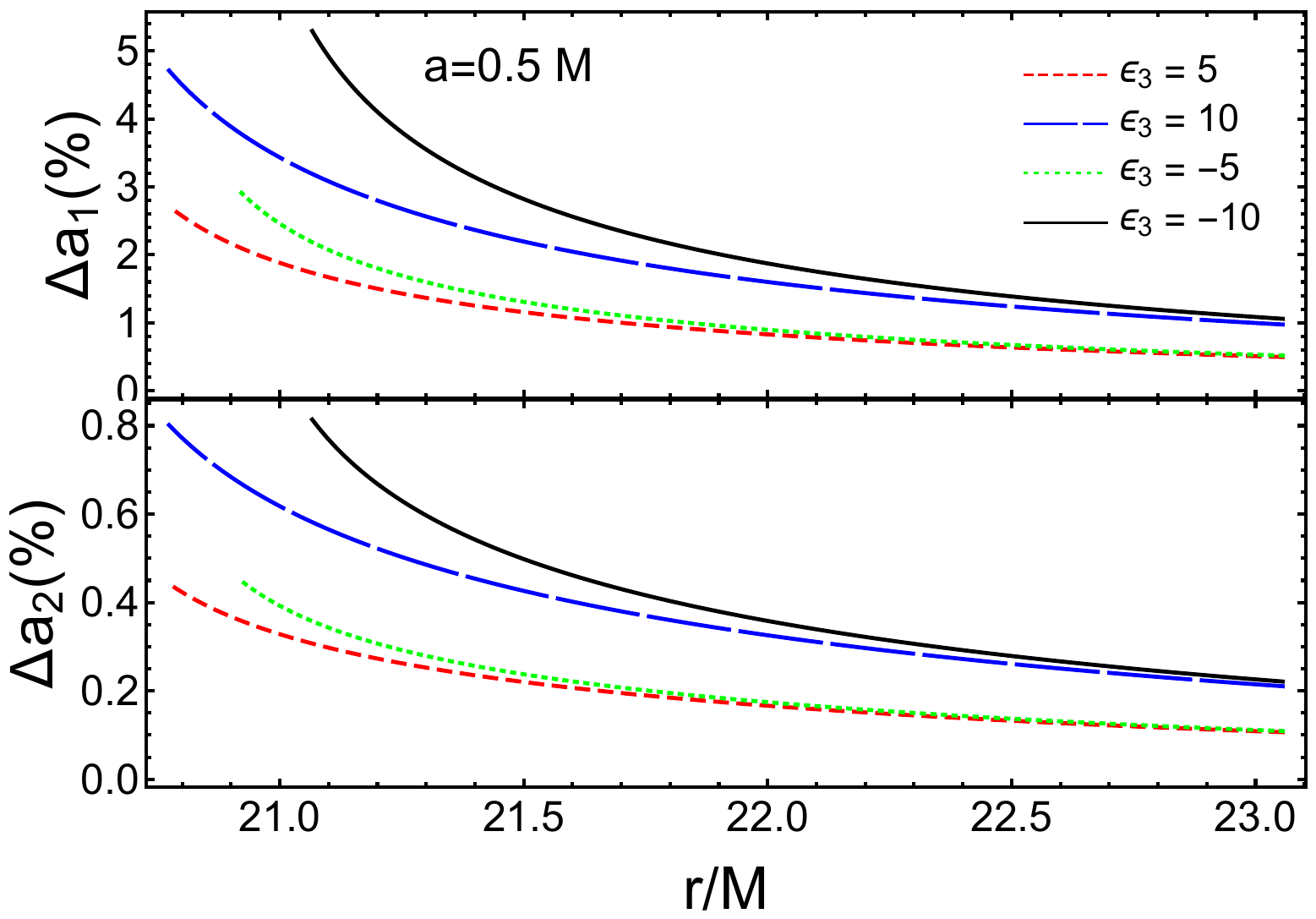}
\caption{We show the difference Eq.~\eqref{deltaa}, between the axes computed 
for $\epsilon_3=(\pm5,\pm10)$ and $\epsilon_3=0$, for $a/M=0.5$, as a function of the
orbital distance divided by the total mass of the system.}
\label{fig6}
\end{figure}
\item In Table~\ref{tableresults}, we show, for each binary configuration considered, 
the critical distances at which the WD fills its Roche lobe and the related values 
of the axes normalized to the star radius at spherical equilibrium, $\bar{a}_i=a_i/R_\tn{WD}$.  
We note that all the simulations end around $r_\tn{tide}\sim 20M_\tn{tot}$. This can be 
explained looking at the behavior of the gravito-electric tidal tensor in Fig.~\ref{fig1}, which shows 
that the effect of the deformation parameter $\epsilon_3$ plays a crucial role only for distances 
$r\sim 10M_\tn{tot}$. Finally, as pointed out in the previous section, values of $\epsilon_3$ 
smaller than zero increase the strength of the tidal field and make the star fill 
its Roche lobe at  larger distances.

It is worth remarking that the orbital radius $r_\tn{tide}$ does not identify the WD tidal 
disruption but the distance at which the star begins to lose mass towards the companion 
object. We expect therefore that the orbital evolution after this critical point, and the 
dynamics of the accreting flow, will be significantly affected by the Kerr metric 
modifications. However, such effects can be tracked only by means of numerical codes. 
This will be the subject of further investigations, in which we will implement 
the theoretical results developed in the previous section into fully relativistic numerical simulations.
\begin{table}[ht]
\centering
\begin{tabular}{cccccc}
\hline
\hline
a & $\epsilon_3$ & $r_\tn{tide}/M$ & $\bar{a}_1$ & $\bar{a}_2$ & $\bar{a}_3$\\
\hline
0.5 & 0 & 20.773 & 1.306 & 0.919 & 0.947\\
0.5 & 5 & 20.621 & 1.296 & 0.919 & 0.948\\
0.5 & 10 & 20.464 & 1.287 & 0.920 & 0.949\\
0.5 & -5 & 20.922 & 1.315 & 0.919 & 0.946\\
0.5 & -10 & 21.066 & 1.323 & 0.919 & 0.945\\
 \hline
0.8 & 0 & 20.738 & 1.303 & 0.919 & 0.947\\
0.8 & 5 & 20.584 & 1.294 & 0.919 & 0.949\\
0.8 & 10 & 20.426 & 1.285 & 0.919 & 0.950\\
0.8 & -5 & 20.888 & 1.312 & 0.919 & 0.946\\
0.8 & -10 & 21.033 & 1.321 & 0.918 & 0.945\\
\hline
\hline
\end{tabular}
\caption{For each binary configuration considered, identified by the BH spin $a/M$ and 
the deformation parameter $\epsilon_3$, we show the critical orbital distance at which 
the simulation ends and the maximum deformation of the WD axes normalized to 
its radius at spherical equilibrium $\bar{a}_i=a_i/R_\tn{WD}$.}
\label{tableresults} 
\end{table}

\item
We have followed the same procedure described above for $a/M=0.8$, finding again 
differences at the most of $\Delta a_{1}\sim5\%$ for $\epsilon_3=\pm10$. As already noted 
therefore, the effect of the BH spin does not change the evolutionary picture. 
This can be easily understood looking at Eqs.~\eqref{E110}-\eqref{E130}: even 
in the standard Kerr case, the spin starts to provide a significant contribution only for 
distances $r<10M$.
\end{itemize}


\section{Conclusions}\label{sec:concl}
Current and future observations in the electromagnetic and gravitational spectrum will 
allow us to map the space-time around supermassive BHs and to study the properties 
of the strong gravitational field in their surroundings. In particular, these experiments will 
shed new light on the validity of the no-hair theorem, for which astrophysical BHs in General 
Relativity belong to the Kerr family, and are described only by their mass and spin. In this 
scenario, the quest for model-independent tests of gravity which make use of the incoming 
flood of data is more needed than ever. Several efforts have been devoted to pursuing this goal. 
Among the model proposed in literature, we have considered the JP metric, which 
parametrizes the deviations from the Kerr geometry through a set of free parameters, 
to be constrained by experiments. 

In this work, we focused our attention on the description of tidal effects produced by rotating 
BHs. We computed the analytic expression for the gravito-magnetic and gravito-electric tidal 
tensors, which completely describe the quadrupolar nature of the tidal field. We consider the 
case of equatorial geodesics, expanding all quantities at the linear order in the parameter 
$\epsilon_3$ (or equivalently $h$), which identifies the deviations from the Kerr metric. 
Comparing our results with those obtained in the pure GR case, we have found discrepancies 
both for ${\cal E}$ and ${\cal H}$ which can be as high as $\sim10\%$ even for large distances, 
$r>10M$. These differences seems also to 
weakly depend on the BH angular momentum. 

We have implemented our results into a semianalytic approach, called the affine model, 
to simulate the encounter of BH-WD systems, following the orbital evolution until 
the star fill its Roche lobe. Analyzing a representative set of binary configurations, 
we have found that the tidal deformations of the WD can be up to $5\%$ different 
between the pure GR and the alternative scenario, for $\epsilon_3=\pm10$, 
even at orbital separation $r\sim 20 M$. Therefore, we expect that the matter flow onto the 
BH, and the possible formation of an accreting disk, would be significantly affected 
by the strong-field correction induced by the JP metric. 

Assessing the features and the detectability of such processes will be a matter of future 
investigations, in which we will implement our theoretical results into fully relativistic 
numerical simulations.


\acknowledgments
It is a pleasure to thank Cosimo Bambi for having carefully read the manuscript and for 
his useful comments. P.L. is supported by the NSF Grants No. 1505824 and No. 1333360. 


\appendix

\section{The parallel transported tetrad in the JP space-time}\label{AppBasis}
In this section, we show the form of the basis vectors $\lambda^{\mu}{_{(\alpha)}}$ 
parallel propagated along a circular geodesic in the JP space-time, at the linear 
order in the parameter $h=\epsilon_3M^3/r^3$,

\begin{widetext}
\begin{subequations}
\begin{equation}
\lambda^{0}{_{(0)}}=\frac{1+a\omega_\tn{k}}{N}-\frac{h}{N^34r^2}\left[3a^2+5r^2+6a(a^2+r^2)\omega_\tn{k}-4r^2(4a^2+3r^2)\omega^2_\tn{K}+12ar^4\omega_\tn{k}^3\right]\ ,
\end{equation}
\begin{equation}
\lambda^{1}{_{(0)}}=0\ ,
\end{equation}
\begin{equation}
\lambda^{2}{_{(0)}}=0\ ,
\end{equation}
\begin{equation}
\lambda^{3}{_{(0)}}=\frac{\omega_\tn{k}}{N}-\frac{h}{4r^2\omega_\tn{k}N^3}\left[6a^2\omega_\tn{k}^2
+a\omega_\tn{k}(9-16r^2\omega_\tn{k}^2)+3(1-2r^2\omega_\tn{k}^2)^2\right]\ ,
\end{equation}
\end{subequations}
\begin{subequations}
\begin{align}
\tilde{\lambda}^{0}{_{(1)}}=&-\frac{\sqrt{\Delta}\ok}{N}\sin\xi+\frac{h\Delta^{-3/2}}{4r^2N^3\ok}\left[(a^2+r^2)^2(3+9a\ok)+2(a^2+r^2)\ok^2(3a^4-8r^4)-4ar^2\ok^3(a^4+11a^2r^2\right.\nonumber\\
&\left.+9r^4)+2r^4(13r^4-3a^2r^2-18a^4)\ok^4+8ar^6(5a^2+6r^2)\ok^5+4r^8(5a^2-3r^2)\ok^6-24ar^{10}\ok^7\right]\sin\xi\ ,
\end{align}
\begin{equation}
\tilde{\lambda}^{1}{_{(1)}}=\frac{\Delta^{1/2}}{r}\cos\xi-\frac{h}{2\Delta^{1/2}}(r-2M)\cos\xi\ ,
\end{equation}
\begin{equation}
\tilde{\lambda}^{2}{_{(1)}}=0\ ,
\end{equation}
\begin{align}
\tilde{\lambda}^{3}{_{(1)}}=&-\frac{1+\ok(a-2r^2\ok)}{N\Delta^{1/2}}\sin\xi+\frac{h}{4r^2N^3\Delta^{3/2}}\left[
6a^5\ok+a^4(3-4r^2\ok^2)+18a^3r\ok (r-2M)+3r(r-2M)^3\right.\nonumber\\
&\left.+4a^2r^2(2-9r^2\ok^2+10r^4\ok^4)+a(6r^4\ok-22r^6\ok^3+20r^8\ok^5)\right]\sin\xi\ ,
\end{align}
\end{subequations}
\begin{subequations}
\begin{align}
\lambda^{0}{_{(2)}}=&0\ ,\\
\lambda^{1}{_{(2)}}=&0\ ,\\
\lambda^{3}{_{(2)}}=&0\ ,\\
\lambda^{2}{_{(2)}}=&\frac{1}{r}\ ,
\end{align}
\end{subequations}
\begin{subequations}
\begin{align}
\tilde{\lambda}^{0}{_{(3)}}=&\frac{\sqrt{\Delta}\ok}{N}\cos\xi-\frac{h\Delta^{-3/2}}{4r^2N^3\ok}\left[(a^2+r^2)^2(3+9a\ok)+2(a^2+r^2)\ok^2(3a^4-8r^4)-4ar^2\ok^3(a^4+11a^2r^2\right.\nonumber\\
&\left.+9r^4)+2r^4(13r^4-3a^2r^2-18a^4)\ok^4+8ar^6(5a^2+6r^2)\ok^5+4r^8(5a^2-3r^2)\ok^6-24ar^{10}\ok^7\right]\cos\xi\ ,
\end{align}
\begin{equation}
\tilde{\lambda}^{1}{_{(3)}}=-\frac{\Delta^{1/2}}{r}\sin\xi-\frac{h}{2\Delta^{1/2}}(r-2M)\sin\xi\ ,
\end{equation}
\begin{equation}
\tilde{\lambda}^{2}{_{(3)}}=0\ ,
\end{equation}
\begin{align}
\tilde{\lambda}^{3}{_{(3)}}=&\frac{1+\ok(a-2r^2\ok)}{N\Delta^{1/2}}\cos\xi-\frac{h}{4r^2N^3\Delta^{3/2}}\left[
6a^5\ok+a^4(3-4r^2\ok^2)+18a^3r\ok (r-2M)+3r(r-2M)^3\right.\nonumber\\
&\left.+4a^2r^2(2-9r^2\ok^2+10r^4\ok^4)+a(6r^4\ok-22r^6\ok^3+20r^8\ok^5)\right]\cos\xi\ .
\end{align}
\end{subequations}
\end{widetext}


\bibliography{bibnote}
\bibliographystyle{apsrev}

\end{document}